\documentclass{epl}

\title{Solid domains in lipid vesicles and scars}
\author{Y. Chushak\inst{1,2} \and A. Travesset\inst{1,2}}
\institute{
  \inst{1} Physics Department, Iowa State University, Ames, IA, 50011, USA\\
  \inst{2} Ames Lab, Ames, IA, 50011, USA}


\begin{document}

\maketitle

\begin{abstract}

The free energy of a crystalline domain coexisting with a liquid
phase on a spherical vesicle may be approximated by an elastic or
stretching energy and a line tension term. The stretching energy
generally grows as the area of the domain, while the line tension
term grows with its perimeter. We show that if the crystalline
domain contains defect arrays consisting of finite length grain
boundaries of dislocations (scars) the stretching energy grows
linearly with a characteristic length of the crystalline domain.
We show that this result is critical to understand the existence
of solid domains in lipid-bilayers in the strongly segregated two
phase region even for small relative area coverages. The domains
evolve from caps to stripes that become thinner as the line
tension is decreased. We also discuss the implications of the
results for other experimental systems and for the general problem
that consists in finding the ground state of a very large number
of particles constrained to move on a fixed geometry and
interacting with an isotropic potential.

\end{abstract}

\section{Introduction}

Spherical crystals appear in many different contexts. Recent
examples include colloids on oil-water interfaces (Colloidosomes)
\cite{Science:03}, micropatterning of spherical particles
\cite{MaIKo:05} relevant for photonic crystals or Clathrin cages,
responsible for the vesicular transport of cargo in cells
\cite{Alberts} (see \cite{KoKrG:03} for a detailed theoretical
study). Crystalline domains covering a fraction of the sphere are
also of experimental interest. In the context of lipid rafts
\cite{SiVaz:04}, confocal fluorescence microscopy studies have
revealed the coexistence of fluid and solid domains on giant
unilamellar vesicles made of lipid mixtures. The shapes of these
solid domains include stripes of different widths and orientations
\cite{KSWFe:99,FeBu:01,SKSch:03} as well as spherical caps
\cite{VeKel:05}. Very recently, solid domains in the gel phase,
clearly displaying stripes and circular domains have also been
investigated by Poon and collaborators \cite{Gordon:05}.

The structure of a crystalline domain on a sphere is just one
example of the more general problem of the interplay between
crystalline order and geometry, which can be understood from two
limiting cases: Crystalline order wins over geometry and the
crystal consists of faceted domains with isolated buckling at
topological defects \cite{SeNel:88,LoBWi:96} (The elastic energy
of a faceted icosahedron maybe computed from \cite{LoBWi:96}) or
geometry is the victor, and the crystal is able to adapt to a
smooth geometry by a proliferation of a well defined array of
topological defects (scars) \cite{BNTr:00}. The intermediate
regimes, which are characterized by the ratio of the bending
rigidity to the product of the Young modulus with the radius
square of the crystal, are of great interest and have been
described in \cite{KoKrG:03,LMNel:03}. For solid lipid domains,
the faceting scenario has been advocated in \cite{LIDim:03}.
Schneider and Gompper \cite{SchGo:95}, however, have pointed out
that the domains observed in \cite{KSWFe:99,FeBu:01,SKSch:03} do
not appear consistent with faceted crystalline domains, and have
discussed the alternative possibility where crystalline domains
remain curved. In that case, the shape of the domains follows from
the competition between the stretching energy and the solid-liquid
line tension. The free energy is
\begin{equation}\label{SG}
    E=E_s+\gamma p  \ ,
\end{equation}
where $E_s$ is the stretching or elastic energy, $\gamma$ the
solid-liquid line tension and $p$ the perimeter of the
liquid-solid boundary. The analysis in \cite{SchGo:95} is an
expansion that works better for small area coverages. Large
domains, which are observed in some experiments \cite{KSWFe:99},
pose some problems, as it has been shown that because of the
finite gaussian curvature spanned by the curved
domains\cite{BNTr:00}, local strains build up, which inevitable
lead to the proliferation of topological defects.

In this paper, we show that scars are necessary to understand the
shapes and structure of large crystalline domains on spherical
vesicles. We determine its structure from general results
\cite{Trav_a:05} derived for arbitrary geometries and show that
with the inclusion of the scars, the stretching energy is
drastically reduced. We believe faceting is unfavorable (to a
larger degree, since experimental results do show small shape
deformations in some cases\cite{KSWFe:99}) because although the
stretching energy of a faceted crystal may only grow
logarithmically with its area \cite{SeNel:88}(such as in the case
of a buckled conical crystal) it implies a much larger line
tension with the coexisting liquid domain. We further assume that
the bending rigidity is the same for both solid and liquid
domains, which is usually a reasonable approximation. We will be
considering a fixed geometry (the sphere) and therefore the
bending rigidity will not play a role in determining the shapes
and structures of the domains.

\section{Stretching energy of spherical caps}

\begin{figure}
\twofigures[height=1.8 in]{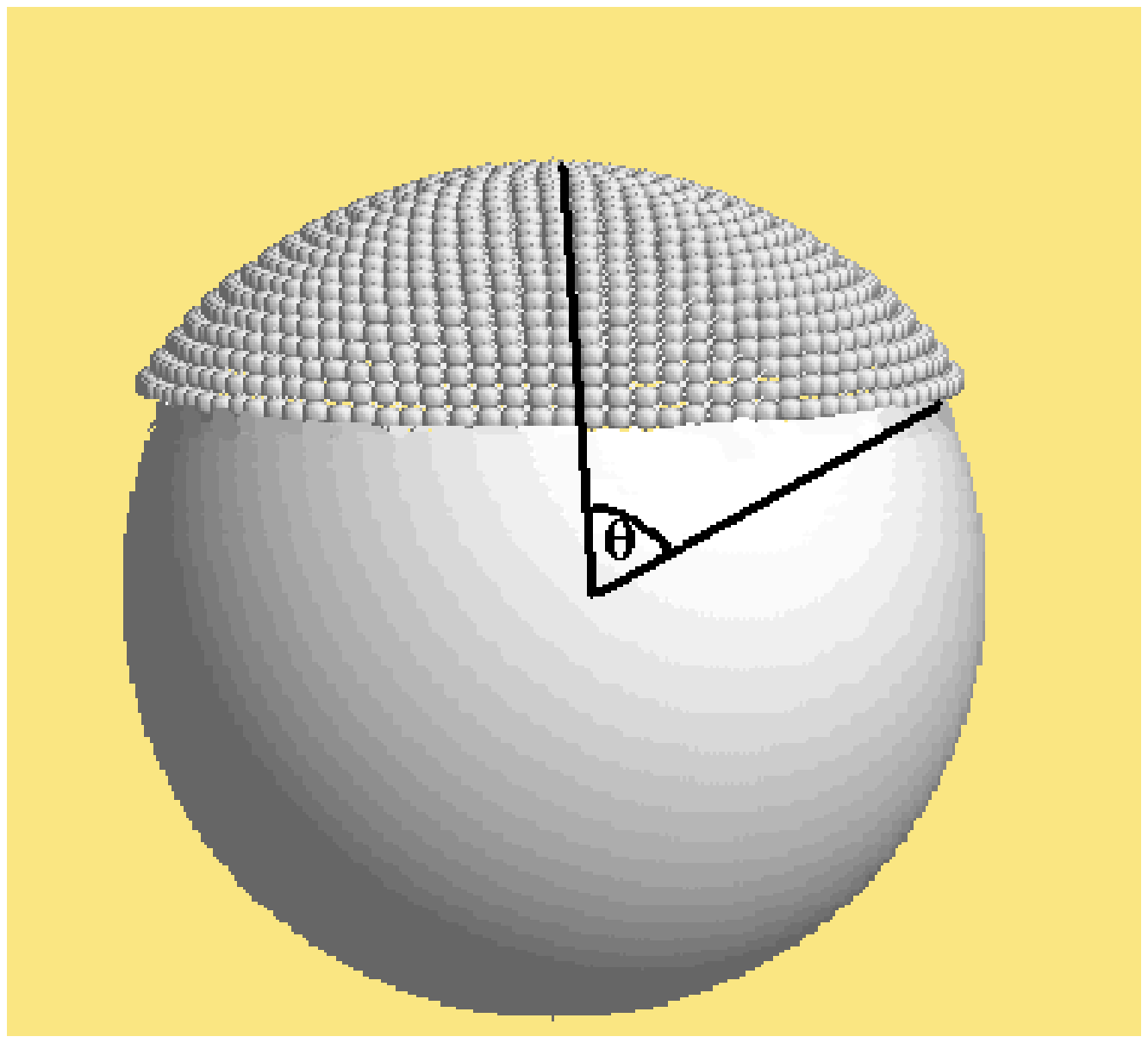}{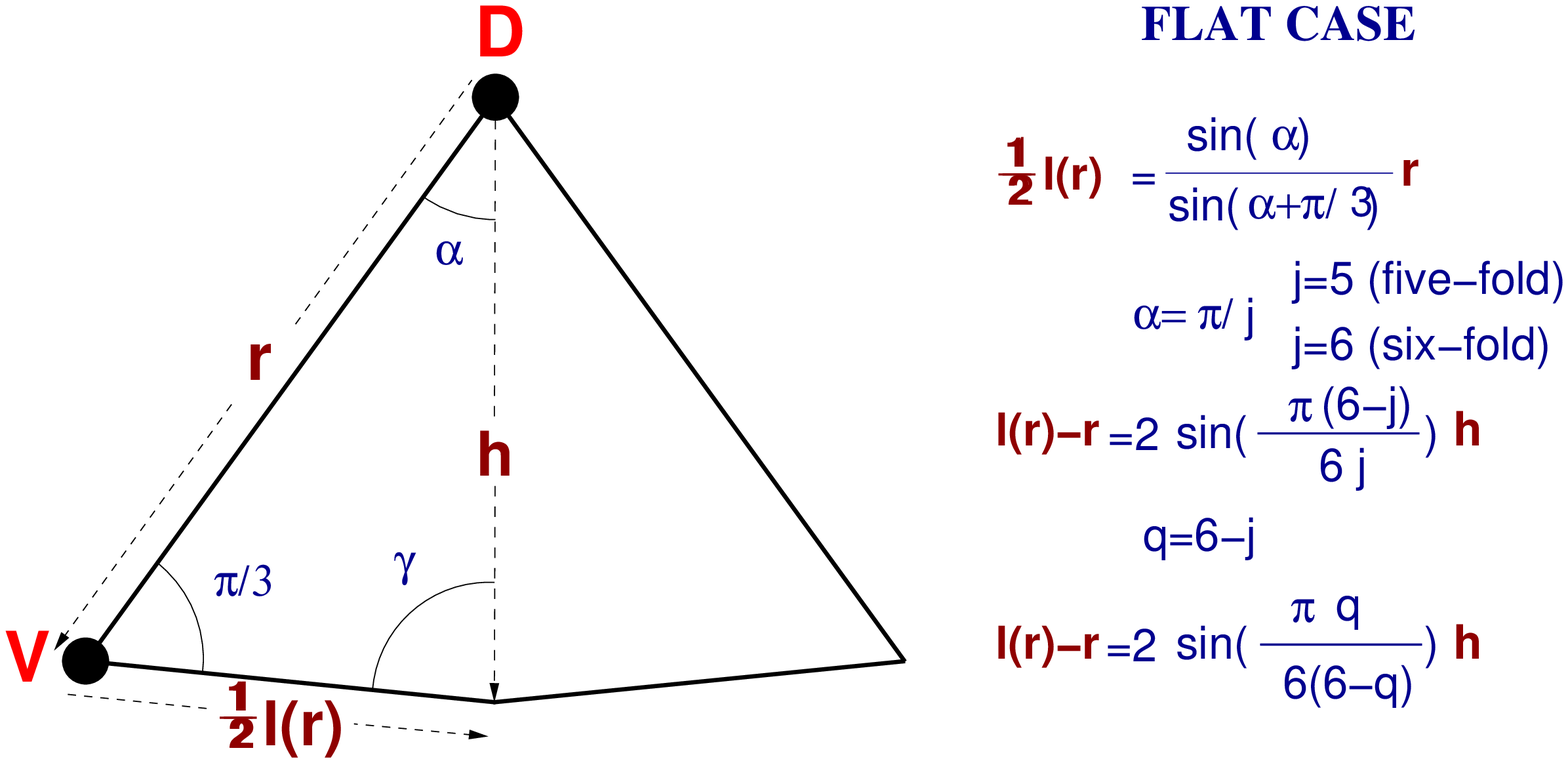}
\caption{Definition of the subtended angle
$\theta$.}\label{fig_theta} \caption{Geometrical argument to
compute the function $l(r)$ on a plane, which is necessary to
obtain the location of the dislocation forming the scars. The same
argument can be generalized to any geometry.}\label{fig__arg}
\end{figure}

We first consider the case of an spherical cap on a sphere of
radius $R$, as shown in fig.~\ref{fig_theta}. The elastic
stretching energy is
\begin{equation}\label{Landau}
E_s=\int d^2 {\vec r}(\mu u^2_{\alpha
  \beta}+\frac{\lambda}{2}(u_{\alpha \beta})^2),
\end{equation}
where $\lambda,\mu$ are the Lame coefficients and $u_{\alpha
\beta}$ is the strain tensor. Upon minimizing the above
expression, the stretching energy of a crystalline domain with
radius $r_a$ becomes
\begin{equation}\label{str_en}
E_s=K_0 r_a^2 F(\theta)
\end{equation}
where $K_0$ is the Young modulus and $\theta=r_a/R$ is half of the
total angle subtended by the cap (see fig.~\ref{fig_theta}). The
angle $\theta$ is related to the relative area covered by the
domain
\begin{equation}\label{Area_covered}
    \frac{A}{A_0}=\sin^2(\frac{\theta}{2}) \ ,
\end{equation}
where $A_0=4\pi R^2$ is the total area of the sphere. The function
$F(x)$ in eq.~\ref{str_en} depends on the actual defect
distribution within the cap. At small values of $x$ for both a
defect free cap or one with a single disclination at its center,
the result is \cite{SchGo:95}
\begin{equation}\label{str_en_a}
F(x) \sim \left\{ \begin{array}{cc} x^4 & \mbox{no
defects} \\
 1-\frac{3}{2} x^2 & \mbox{five-fold disclination at the center of the domain}\end{array}\right.
\end{equation}
for larger values of $x$ the function $F(x)$ might be computed
from the results in \cite{BNTr:00}. We are interested in the
regime where $\theta$ is not necessarily small, and in that case
the stretching energy grows quadratically with the radius of the
cap. For reasonable values of the elastic constants and the line
tensions, the stretching energy will only allow for small domains
($\theta<<1$).

\begin{figure}
\onefigure[width=3.5 in]{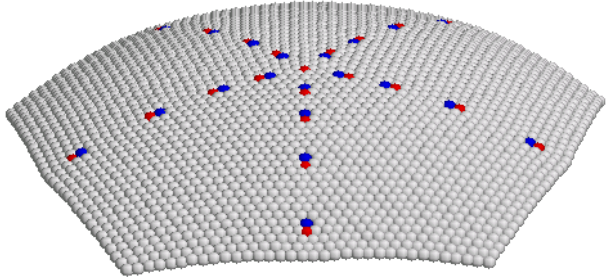}\caption{Example of a typical
scar on a spherical cap. The center of the domain is a five-fold
$q=1$, and 5 grain boundaries each containing four dislocations,
with increasing distance within the grain, form the scar. Results
are for $\theta=0.6 \approx 34.4^o$ ($A/A_0 \approx 0.9$) and
$L=30$.} \label{fig_scar}
\end{figure}

As already pointed out, in curved geometries the energy maybe
reduced by the inclusion of scars. In \cite{Trav_a:05} the precise
structure of these scars was determined from analytical arguments.
It was shown that the stretching energy of a crystalline domain
consisting of a spherical cap, once scars are included, for any
value of $\theta$, grows only linearly with the domain radius
\begin{equation}\label{str_en_m}
E_s \sim K_0 a r_a \ ,
\end{equation}
where $a$ is the lattice constant of the crystal. The actual
structure of the scars is obtained from the following; First, the
cap is divided into $j$ identical triangular wedges of angle
$2\alpha=2 \pi/j$, where $j=5,6$ depending on whether the center
of the cap is a five or six-fold vertex. The disclination charge,
defined as $q=6-j$ is $q=1$ for a five-fold and $q=0$ for a
six-fold vertex. The function $l(r)$, which measures the failure
to close the wedge with equilateral triangles, is computed from
the geometric construction described in fig.~\ref{fig__arg} for
the geometry of interest (in our case the sphere). The location of
the dislocations defining the scars is obtained from the equation
\begin{equation}\label{cond}
    l(r)-r= n a
\end{equation}
where $n=\pm1,\pm2,\pm3,\cdots.$ Additional details, including the
explicit form for the function $l(r)$ as well as the proof that
such construction leads to a stretching energy with the form of
eq.~\ref{str_en_m} can be found in \cite{Trav_a:05}, but it can be
more intuitively understood from noticing that the planar result
that leads to eq.~\ref{str_en_m} is a consequence of the triangles
on the lattice being very close to equi-lateral. On a general
geometry, we can achieve a similar close to equilateral
configuration by noticing that Gaussian curvature modifies the
area of the wedge at a given radial distance from the center, and
therefore, the equi-spacing of dislocations that follows for
planar geometries needs to be distorted according to the Gaussian
curvature, which is eq.~\ref{cond}. We now show by an explicit
numerical minimization that the scars defined by eq.~\ref{cond}
have an energy defined by eq.~\ref{str_en_m}. An example of a
typical configuration with scars is shown in fig.~\ref{fig_scar}.

\section{Numerical Results}

We discretize the elastic energy eq.~\ref{Landau} by considering a
discrete triangular lattice with the topology of a disk and open
boundary conditions. The discrete stretching energy is
\begin{equation}\label{Eq1}
E_1(\varepsilon)
=\frac{\varepsilon}{2}\sum_{(b,c)}(|\vec{r}_{bc}|-a)^2
\end{equation}
where $r_{bc}$ is the nearest-neighbor distance. The coupling
$\varepsilon$ is related to the Lame coefficients according to
$\mu={\sqrt{3}\varepsilon}/{4}$ and
$\lambda={\sqrt{3}\varepsilon}/{4}$ \cite{SeNel:88}. We now
consider the cap as made of $L$ monomers along the radial distance
from its center, so that $r_a=L a$, we write eq.~\ref{str_en} as
\begin{equation}\label{str_en_num}
E_s=K_0 a^2 L^2 F(\theta)
\end{equation}

\begin{figure}
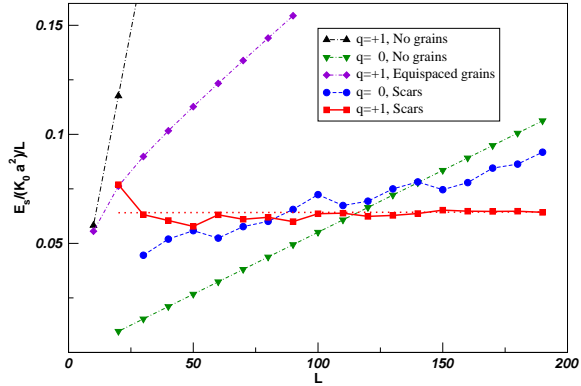

\onefigure[width=3in]{Energy_theta_0p6_all.eps}\caption{Stretching
energy in units of $K_0 a^2$ divided by $L$ as a function of the
lattice size $L$. The label $q=1,0$ represents whether there is a
plus-disclination or a regular vertex at the center of the
crystal. Equi-spaced configuration is the configuration that
minimizes the plane for $q=+1$. The scars are predicted from
eq.~\ref{cond} as explained in the text. The dotted line is a fit
to $q=+1$ scars. Results are for $\theta=0.6$ ($A/A_0 \approx
0.9$), and $\mu=\lambda=0.4333$.} \label{fig_theta0p6}
\end{figure}

Let us recall the results for the plane \cite{Trav:03}. Obviously,
a perfect triangular lattice with no defects has the lowest
stretching energy (zero, in our case). If, however, a planar disk
has a disclination at its center, then the scars consist of grain
boundaries with equi-spaced dislocations (whose spacing is
calculated in fig.~\ref{fig__arg}) and the energy grows linearly
with $L$, a result that can also be proven analytically
\cite{Trav:03}. Let us now explore the results at finite
curvature. It is practical to consider a fixed aperture angle and
consider the stretching energy (in units of $K_0 a^2$) divided by
the dimensionless number $L$, so that if the stretching energy
behaves linearly with $L$ (as in eq.~\ref{str_en_m}) it will show
up as a constant as a function of $L$. Results for $\theta=0.6
\approx 34.4^o$(${A}/{A_0}\approx 0.9$) are shown in
fig.~\ref{fig_theta0p6} (The actual configuration for $L=30$ is
shown in fig.~\ref{fig_scar}). For configurations consisting of
either a perfect lattice (defect free), a single disclination with
no defects, or a central disclination with the equi-spaced radial
grain boundaries that minimize the energy in the planar case, the
stretching energy grows with the expected quadratic behavior
eq.~\ref{str_en}, as clearly evidenced by the linear growth with
$L$ in the plot. With a five fold at the center and the scars
defined by eq.~\ref{cond}, however, the plot of the stretching
energy is horizontal, implying that the energy grows linearly with
$L$ (eq.~\ref{str_en_m}). The largest system investigated $L=200$,
consists of the order of $10^5$ monomers and contains $\sim 120$
dislocations. If the center of the cap has a six-fold vertex, the
stretching energy does show a slight slope. This is due that the
theoretical argument eq.~\ref{cond} places a large number of
dislocations close to the boundaries. In other words, the
predicted theoretical configuration is very difficult to realize
in practice.

\begin{figure}
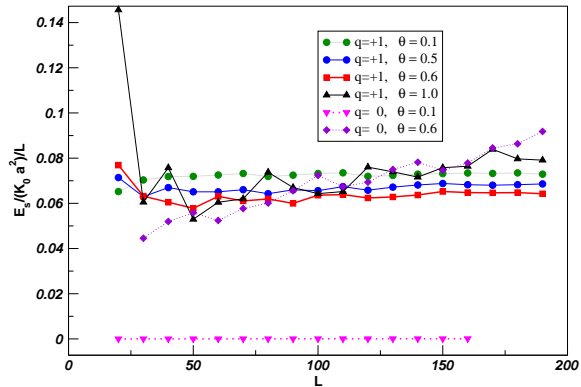

\onefigure[width=3in]{Energy_comp.eps}\caption{Stretching energy
in units of $K_0 a^2$ divided by $L$ as a function of the lattice
size $L$ for different aperture angles $\theta$ and the variable
spacing is the energy for the dislocations separation predicted by
eq.~\ref{cond} as explained in the text. Results are for
$\mu=\lambda=0.4333$.} \label{fig_all_theta}
\end{figure}

Results for aperture angles within the range $\theta=0.1\approx
6^o$ (${A}/{A_0}\approx 3\cdot10^{-3}$) to $\theta=1.0\approx
60^o$ (${A}/{A_0}\approx 0.25$) are shown in
fig.~\ref{fig_all_theta}. The numerical results show remarkable
agreement with the expected behavior eq.~\ref{str_en_m}. Even for
small area fractions (${A}/{A_0}=0.05$) a plus disclination and
scars are favored, and that for ${A}/{A_0}> 0.04$ the coefficient
of the stretching energy eq.~\ref{str_en_m} is roughly independent
of $\theta$. Only for the largest value of $\theta$ studied
($\theta=1$), a slight slope is observed in the figure. We believe
that this is a consequence that by keeping open boundary
conditions, the aperture angle $\theta$ is modified from its
original value.

There are a few technical aspects that we describe here. The
dislocations forming the scars can only be placed at positions
defined by lattice points, that is, positions that are integers of
the lattice constant. The positions predicted by eq.~\ref{cond}
are in general not integers so the result was rounded, not just to
the nearest integer, but to the first odd integer (this choice
respects better the structure of the underlying lattice). The
defects were placed to the desired positions by using the software
developed in \cite{Trav:03}.

\section{Discussion and Conclusions}

We have shown explicitly that the stretching energy of a spherical
cap on a sphere grows linearly with its radius when scars are
introduced following eq.~\ref{cond}. A typical scar on a spherical
cap is shown in fig.~\ref{fig_scar}. We now consider crystalline
domains different than the cap. A ring can be considered as the
region defined by two spherical caps of radius $r_2,r_1$, with
$r_2 > r_1$. The width of the ring is $d$, where $d=r_2-r_1$. We
further assume that the ring is thin $d<< r_1$ but not microscopic
$d>> a$. The stretching energy of a ring containing scars can be
computed as the difference of the stretching energy of the two
caps, $E_{st}\sim E_{st}(r_2)-E_{st}(r_1) \sim  K_0 a d$ (elastic
energy alone tries to avoid Gaussian curvature). Let us now
compare the stretching energy of a ring with the stretching energy
of a cap, both covering the same area. It is found,
\begin{equation}\label{ribbon_cap}
  E_{st}^{ring} \sim  \sqrt{\frac{d}{r_1}} E_{st}^{Cap}   \ ,
\end{equation}
since ${d}/{r_1} << 1$, the stretching energy favors rings over
caps (both containing scars).

Let us examine now the free energy eq.~\ref{SG} containing both
stretching and line tension energy terms. For a cap of radius
$r_a$ and area coverage ${A}/{A_0} > 0.04$, the free energy is
\begin{equation}\label{new_energy}
E \approx {0.07}{K_0 a r_a}+ 2\pi \gamma r_a \ ,
\end{equation}
where the pre-factor, which we do not claim to be universal as it
may depend on the Poisson ratio, is obtained from
fig.~\ref{fig_all_theta}. If
\begin{equation}\label{high_lambda}
\gamma >> \frac{0.07}{2\pi} {K_0 a} \ ,
\end{equation}
the line tension dominates and a single cap with a disclination at
its center and scars will be favored. If the fraction of solid
phase is small (${A}/{A_0}<<0.04$), the cap will consist of a
regular vertex and scars. As the line tension is decreased below
${0.07K_0 a}/{2\pi}$, the stretching energy begins to dominate and
the system will evolve into rings (or stripes), becoming thinner
as the surface tension is decreased. A typical radius for a
unilamellar vesicle is $R\sim 30 \mu m$. In this case, a domain as
small as $1 \mu m$ should contain a significant fraction of
defects. Typical molecular area of a phospholipid is $\sim 60
\AA^2$ (is actually lower in the crystalline phase). For an
aperture angle $\theta=0.6$ this gives $L\sim 3500$. As it is
clear from fig.~\ref{fig_all_theta}, this value of $L$ is deep
into the regime were the stretching energy is linear in $L$. As a
concrete example, in \cite{SchGo:95} the parameters for a
DLPC/DMPC mixture are estimated as $\lambda\approx k_B T/a$, $K_0
\approx 500 k_B T/a^2$. In this case we obtain ${0.07K_0 a}/{2\pi}
\approx 5 k_B T/a > k_B T/a \approx \gamma$. That is, the
stretching energy dominates and stripes are favored. The effect of
gradually adding cholesterol to a DLPC/DHDP mixture amounts to a
reduction of the solid-liquid line tension, and as a consequence,
the stripes should become thinner. All these observations are in
agreement with the experimental results \cite{KSWFe:99}.

In this paper, we have ignored the effect of thermal fluctuations
on the scars. This problem has been experimentally investigated
with colloidosomes in \cite{LBMNB:05}, where it has been shown
that dislocations fluctuate (by glide) but on average remain in
their equilibrium positions. We expect similar results to apply in
general, but the discussion of this point is beyond the scope of
this paper.

The results presented in this paper have a broader interest beyond
solid domains in lipid mixtures. Our results directly apply to
other experimental systems, such as large colloidosomes
\cite{Science:03} and the spherical arrays considered in
\cite{MaIKo:05}. The numerical evidence summarized by
fig.~\ref{fig_all_theta} and fig.~\ref{fig_theta0p6} provide a
numerical verification to the theoretical ansatz given in
\cite{Trav_a:05} for the ground state of a large number of
particles constrained on a given geometry and interacting with an
isotropic potential. This is a problem of vibrant activity in the
field of potential theory in mathematics, where recent rigorous
results have been proved \cite{HarSa:05}. Our results provide an
explicit construction for the ground state, and allows
straight-forward generalizations to other geometries. Progress for
the torus and negative curvature surfaces will hopefully be
reported soon.

\acknowledgments

We acknowledge many discussions with Mark Bowick, Doug Hardin,
David Nelson, Edward Saff and David Vaknin. AT acknowledges the
Benasque center for sciences where some parts of this paper where
completed. This work has been supported by NSF grant DMR-0426597,
Iowa State Start-up funds and partially supported by DOE under
contract No. W-7405-ENG-82.

\end{document}